# Power Management of Microgrid Integrated with Electric Vehicles in Residential Parking Station

Hojun Jin
The Cho Chun Shik Graduate School of Green Transportation
Korea Advanced Institute of Science and Technology
Daejeon 34141, Republic of Korea
hjjin1995@kaist.ac.kr

Sarvar Hussain Nengroo
The Cho Chun Shik Graduate School of Green Transportation
Korea Advanced Institute of Science and Technology
Daejeon 34141, Republic of Korea
sarvar@kaist.ac.kr

Sangkeum Lee
Environment ICT Research Section
Electronics and Telecommunications Research Institute (ETRI)
Daejeon 34141, Republic of Korea
sangkeum@etri.re.kr

Dongsoo Har[*]
The Cho Chun Shik Graduate School of Green Transportation
Korea Advanced Institute of Science and Technology
Daejeon 34141, Republic of Korea
dshar@kaist.ac.kr

*Abstract*— Lately, increasing number of electric vehicles (EVs) in residential parking station has become an important issue, because excessive number of EVs can destabilize the power system during peak hours with high charging power requested. When the power system of the residential parking station takes the structure of microgrid (MG), power provision for the EVs requires efficient power management scheme. To minimize the maintenance cost of the MG and maintain the grid stability, the MG needs to balance the charging/discharging power of EVs in the parking station. To achieve these goals, this paper proposes a charging/discharging algorithm suitable for the power management of the MG configured with EVs. Multi-objective optimization is taken to MG to minimize the maintenance cost and the grid dependency while maximizing the use of photovoltaic (PV) power and the utilization of EVs as energy storage systems (ESSs). In our approach, to increase the usefulness of discharging power of EVs, the base load and the PV power production are considered together for power management to mitigate imbalances incurred between them. As a result, proposed approach demonstrates superior power management performance as compared to other comparable ones.

I. INTRODUCTION

Microgrid (MG) is typically utilized to improve the flexibility and reliability of power systems. It is typically composed of distributed energy resources (DERs), electrical loads, renewable energy systems (RESs), and energy storage systems (ESSs). Topology of MG can be categorized into island mode and grid-connected mode [1]. In the island mode, the controller balances the load demand with power generation, preventing curtailments of load demand. In a grid-connected mode, the MG is connected to utility grid and thus power management to meet load demand becomes easier. In case of grid-connected mode, MG islanding can occur due to outages of utility grid. A MG is adequately economical and reliable system as it is composed of distributed power generations and aggregated loads [2]. As the fossil fuel problem and environmental concern increase, integrating renewable energy resources with MG becomes essential. In a centralized MG, MG central controller (MGCC) monitors the power generated from DERs and controls the electrical loads to manage the energy balance in the power system. The MGCC reduces the maintenance cost of MG for users as it supplies the power to load demand with its own energy resources when energy prices are high in electricity market. Typically, if the load demand is higher than the power generation, non-critical loads are not allowed to be operated. In [3], electric appliances in nanogrids are classified into shiftable and non-shiftable ones and each shiftable electric appliance is scheduled to balance the load demand and the power generation. The peer-to-peer (P2P) electricity trading is an another option to alleviate the power balance between nanogrids [4]. Nengroo et al.[5] attempted to minimize power consumption by using a hybrid system consisting of photovoltaic (PV) system, battery storage, and utility grid.

Dependency of energy consumption on the fossil fuel increases is a main cause of energy crisis and global warming [6]. For this reason, electric vehicle (EV) becomes the viable alternative of internal combustion engine vehicle as it reduces the greenhouse gas emission. In the MG, the deployment of EVs can be an efficient and economical solution as the ESSs [7]. By properly utilizing EVs as ESSs in power network connected with RESs, the emission of greenhouse gases is significantly reduced [8]. In spite of these advantages, many technical challenges still exist. For example, excessive electricity demand, incurred by simultaneous charging of EVs, can destabilize the power system [9]. Unlike other charging applications such as charging rechargeable sensors [10], [11], charging timing of EVs in MGs should be considered with discharging timing of them in order to achieve efficient power management of MGs. In effect, a new dimension of MG power management is created by considering charging/discharging EVs [12].

Vehicle-to-grid (V2G) technology typically makes the power system more flexible and reliable with intermittent RESs. The V2G configuration allows bi-directional power flow between the utility grid and the battery of EV [13]. In some studies [14], [15], V2G technology is applied to an energy management system with load scheduling in a MG. Based on a flexible control and dispatch strategies, EV batteries can be utilized as ESSs to mitigate peak load in the power system, stabilize the power grid, and mitigate the adverse effect of intermittent RESs [16], [17]. In [18], an optimal power management is applied to the grid-connected office building with joint presence of EVs and PV power source. Timing of charging and discharging EVs is determined by considering daily variation of PV power production. The charging procedure of EVs is regulated to feed power to the utility grid by charging EVs during off peak hours and discharging during high peak hours with predicted PV power production [19], [20]. The EV owners can get profit by selling electricity to the electric utility while their EVs are parked. In [21], the effect of a large number of plug-in hybrid electric vehicles (PHEVs) and battery electric vehicles (BEVs) on the power systems of five Northern European countries is studied. It is shown in [16] that intelligent charging/discharging of EVs significantly maximizes the revenue of wind power system. The integration of PHEVs with MG helps in improving the reliability and flexibility of power management in MG [22]. In [23], the charging/discharging management of EVs in a parking lot is presented to maximize the state of charge (SoC) of each battery of EV. In [24], a power management algorithm has been developed for a grid-connected parking station installed with the PV system. The parking station is aimed at reduction of total electricity cost for charging the PHEVs and mitigation of peak load.

In this paper, we propose a novel approach for power management of MG integrated with EVs in the residential parking station. Proposed power management for MG integrated with the operation of EVs is involved with time varying base load and temporal PV power production, unlike previous studies [25], [26] dealing with power management of parking station without considering base load that is fundamentally important for power management of MG. Power scale of such power systems corresponds to the MG. There are mainly two electricity sources; utility grid and PV system. Each EV is described as one of the electricity sources when it discharges in the parking station. The total electricity load is composed of base load and EV charging load. The base load refers to the existing load demand in MG. To reduce the peak grid power consumption incurred by adding EVs' charging load to base load, EV discharging is utilized by comparing base load and PV power production. In the aspect of the MG operator, the PV power is purchased from the PV system by paying the requisite charges. The purchased PV power can be supplied for EV charging or to the base load. When PV power is insufficient for instantaneous power consumption of MG, additional power can be supplied by the utility grid.

This paper is organized as follows. In Section II, the overall system architecture is explained. Section III describes the proposed power management method for the EV parking station. Section IV provides the simulation results with the proposed power management, and Section V presents the conclusion drawn from the work presented in this paper.

## II. SYSTEM MODEL

### A. Overall System Architecture

This section describes the overall system architecture of the MG consisting of base load, energy resource systems (utility grid and PV system), and EV parking station, as shown in Fig. 1. For the connection with the power network, point of common coupling (PCC) is applied to the system. The utility grid and the PV system can be mainly utilized to serve the load demand in grid-connected mode. The MGCC as communication interface controls the energy flow between the utility grid and the MG, and takes the responsibility to manage the MG efficiently. The load demand consists of base load and EV charging load. Typically, the power system can be significantly unstable with an increasing peak load incurred by simultaneous charging of EVs. Therefore, EV parking station has an important role as an aggregator to manage the EVs' participation and scheduling in the power network.

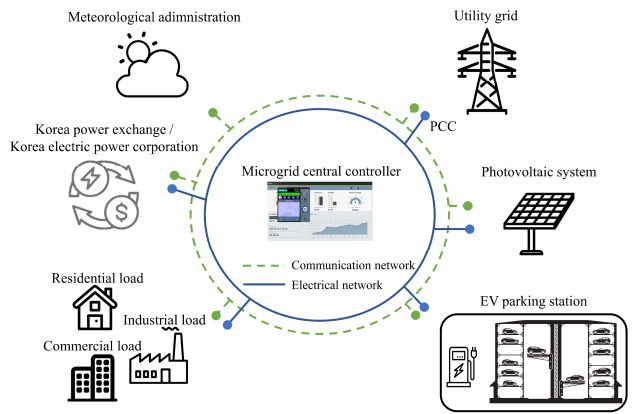

Fig. 1. Overall system architecture.

### B. Multi-EV parking station

In the EV parking station, EVs are used as electrical components in the power network. The energy management system of the parking station typically manages the charging/discharging of EVs for efficient power management of the MG. The information flow of the parking station is described in Fig. 2. To satisfy EVs with different operation conditions, two charging modes $M_1$ and $M_2$ are used. Charging rate for mode $M_1$ is 7kW and charging rate for mode $M_2$ is 19.2kW. In this work, discharging rate for each EV is set to be equal to the charging rate. When an EV is parked, the parking station receives information on EV operation conditions, such as arrival time, departure time, initial SoC, and goal/maximum SoC from the EV. The MGCC receives the information of the base load, system marginal price (SMP) determined by the Korea Power Exchange (KPX), and rate of electricity specified by demand response (DR) program determined by the Korea Electric Power Corporation (KEPCO). Depending on the solar irradiance data obtained from the Korea Meteorological Administration (KMA), maximum PV power production is calculated. In the decision making center, each EV's operation is determined based on the EV owner's information and the comparison of the temporal base load and instantaneous PV power production in the MG.

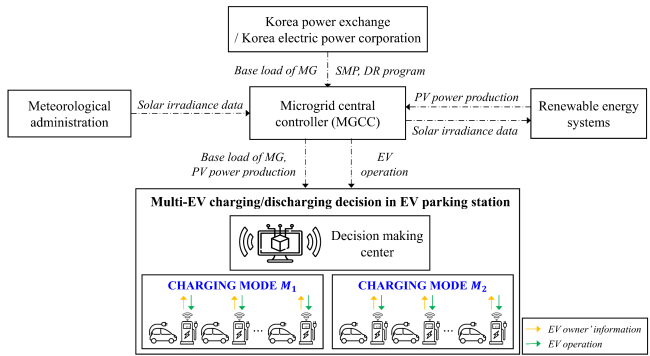

Fig. 2. Information flow in parking station.

*C. Driving Pattern of EVs*

Figure 3 represents the probability of a vehicle when it arrives at home and at workplace. Similar behavior of vehicles has been studied in [27], [28], dealing with parking and driving time. Figure 3 characterizes the daily commuters who arrive at home in the evening (15:00-21:00) and at workplace in the morning (7:20-13:20). The time interval is 15 minutes, thereby producing a total of 96 time slots for a day. As the parking station is located in the residential area, each EV's arrival time to parking station is determined based on the arrival probability at home, and its departure time from parking station is determined based on the arrival probability at workplace. In addition, it is assumed that each EV has a battery with the capacity of 64 kWh and it should be charged up to 80% SoC while being parked. The initial SoC of each EV before charging follows the normal distribution in which the mean and the standard deviation values are 15 and 5, respectively.

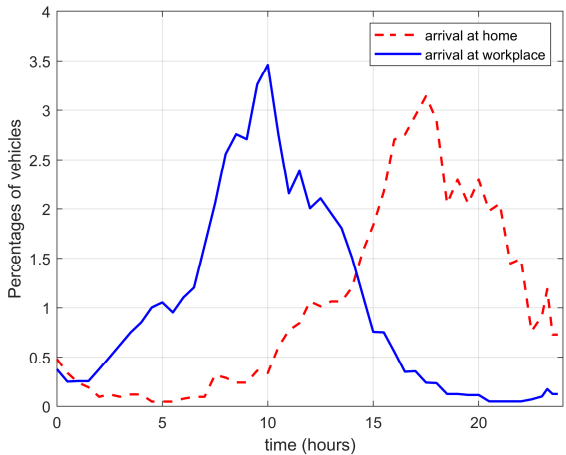

Fig. 3. Distribution of arrival time of vehicles at home and workplace.

*D. Electricity Rate (DR program and SMP)*

For the power management of MG, DR program is utilized as the electricity price of the utility grid [29]. It encourages customers to consume less power during peak hours or to shift the usage of electricity during off-peak hours to flatten the load demand. Based on the DR program, the flexible load, which is EV charging load of the parking station, can be scheduled to reduce the amount of grid power usage and the electricity cost imposed from the utility grid. The electricity rate specified by the DR program is $0.055/kWh during 23:00-09:00, $0.108/kWh during 09:00-10:00, 12:00-13:00, and 17:00-23:00, and $0.179/kWh during 10:00-12:00 and 13:00-17:00, as presented in Fig. 4. In South Korea, the produced PV power can be sold to KEPCO which represents the electric utility. The price of PV power is determined by the SMP, as presented in Fig. 4. It is estimated based on the intersection between the day-ahead electricity load demand curve and the generator bidding supply curve [30]. The electricity cost for purchasing PV power is calculated by multiplying the amount of PV power with SMP.

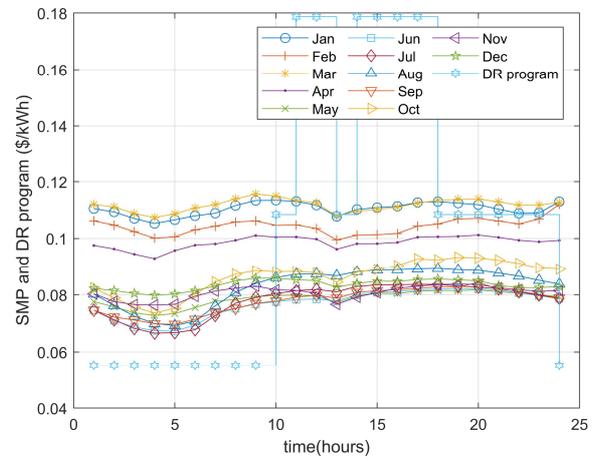

Fig. 4. Electricity rate by DR program and average monthly SMP.

III. POWER MANAGEMENT OF MICROGRID

In this paper, the aim of power management of EVs in residential parking station is to minimize the maintenance cost of the MG in the grid-connected mode. The MGCC receives the information of the base load, PV power production, SMP, and DR program. The information is utilized to determine each EV's operation in every time interval.

*A. Proposed Power Management for EV Parking Station*

For each EV, the switching functions for charging and discharging are defined separately, to reduce the adverse effects such as capacity fades caused by intermittent charging/discharging [31]. Each EV is assigned to the charging/discharging operation based on the comparison of the base load and the PV power production in the MG, as defined in equation (1).

$$\begin{cases} O_{Ch}^{i,t} = +1 \ \ if \ P_{load}^t - P_{PV}^t < P^{flag} \ and \ SoC_{EV}^{i,t} < SoC_{max} \\ O_{Dch}^{i,t} = +1 \ \ if \ P_{load}^t - P_{PV}^t \geq P^{flag} \ and \ SoC_{EV}^{i,t} > SoC_{min} \end{cases}$$

where $O_{Ch}^{i,t}$ and $O_{Dch}^{i,t}$ are the switching functions of the *i*-th EV, which represent charging operation and discharging operation, respectively, and $P_{load}^{t}$ and $P_{PV}^{t}$ are the base load and PV power production at the *t*-th time interval. The $P^{flag}$ is the flag parameter and is used to decide the period that EV charging or discharging is activated, and the $SoC_{EV}^{i,t}$ is the SoC of the *i*-th EV, and $SoC_{min}$ and $SoC_{max}$ are the minimum and maximum allowable SoC of the *i*-th EV. In the proposed method, if the difference between the base load and PV power production is smaller than $P^{flag}$ and EV's SoC is less than $SoC_{max}$, then the $O_{Ch}^{i,t}$ becomes +1 (charge). If it is bigger than $P^{flag}$ and EV's SoC is higher than $SoC_{min}$, then the $O_{Dch}^{i,t}$ becomes +1 (discharge).

### B. Multi-objective optimization for power management of MG

*1) Minimization of electricity cost*

The first objective function aims to minimize the total electricity cost of the MG. The electricity cost is reduced by the scheduling of the EV charging/discharging as follows:

$$OBJ(1) = \pi_{grid}^{t} \times (P_{grid-load}^{t} \times O_{grid-load}^{t} + P_{grid-EV}^{t} \times O_{grid-EV}^{t}) \\ + \pi_{PV}^{t} \times \begin{pmatrix} P_{PV-load}^{t} \times O_{PV-load}^{t} + P_{PV-EV}^{t} \times O_{PV-EV}^{t} \\ - \sum_{i=1}^{N} P_{Dch}^{i} \times O_{Dch}^{i,t} \end{pmatrix} \quad (2)$$

where the $\pi_{grid}^{t}$ and $\pi_{PV}^{t}$ are the rate of electricity from the utility grid and PV system, and follow the DR program and SMP, respectively, and the $P_{grid-load}^{t}$ and $P_{grid-EV}^{t}$ are the electric power delivered to the base load and charging demand from the utility grid, respectively, and the $P_{PV-load}^{t}$ and $P_{PV-EV}^{t}$ are the electric power delivered to the base load and charging demand from the PV system, respectively, and the $O_{grid-load}^{t}$, $O_{grid-EV}^{t}$, $O_{PV-load}^{t}$, and $O_{PV-EV}^{t}$ are the switching functions of the $P_{grid-load}^{t}$, $P_{grid-EV}^{t}$, $P_{PV-load}^{t}$, and $P_{PV-EV}^{t}$, respectively, and the $P_{Ch}^{i}$ and $P_{Dch}^{i}$ are the charging and discharging power of the *i*-th EV, respectively. The charging/discharging power of each EV is assigned depending on the EV owner information. The EVs' discharging power is sold to KEPCO at the price of SMP and mitigates the electricity usage from the utility grid.

*2) Minimization of grid dependency*

The second objective function is considered to minimize grid power consumption for base load and EV charging load. The utilization of RESs becomes more significant as excessive increase of grid power consumption leads to instability in power system. Reduction of the grid power usage can be achieved by minimization as follows:

$$OBJ(2) = P_{grid-load}^{t} \times O_{grid-load}^{t} + P_{grid-EV}^{t} \times O_{grid-EV}^{t} \quad (3)$$

*3) Maximization of the PV power use*

In the MG, the PV power is purchased from PV system at the rate of SMP. When the SMP is higher than the rate of DR program, less PV power can be used to the total load demand. More PV power can be utilized with the following objective function. It includes the PV power used for the base load and EV charging load.

$$OBJ(3) = -P_{PV-load}^{t} \times O_{PV-load}^{t} - P_{PV-EV}^{t} \times O_{PV-EV}^{t} \quad (4)$$

### IV. SIMULATION RESULTS

The capacity of EV parking station is assumed to be 50 EVs. Also, the daily base load and the PV power production of the MG are shown in Fig. 5. In the 'PV-system+proposed-EV-scheduling' method, the power management with the proposed EV scheduling is applied, which includes the PV system and EV discharging. In the 'PV-system+EV-scheduling' method, the power management with a typical EV scheduling is applied, which includes a PV system, but does not include EV discharging. In the 'no-PV-system+no-EV-scheduling' method, the power management and EV scheduling are not applied to the MG, and the PV system does not exist.

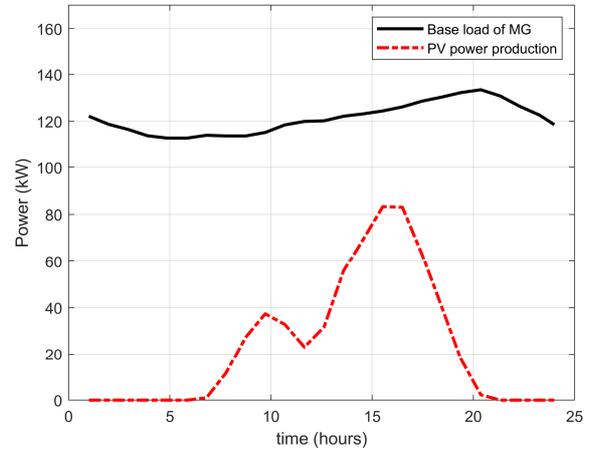

Fig. 5. Variation of base load and PV power production over a day considered for simulations.

Figure 6 describes the grid power consumption depending on the power management scheme. In the 'no-PV-system+no-EV-scheduling' method, the grid power usage is observed to be high without limitation of maximum grid power usage between 0:00 and 5:00, since EVs typically start charging as soon as they are parked. However, less grid power consumption is observed with the 'PV-system+EV-scheduling' method and 'PV-system+proposed-EV-scheduling' method between 0:00 and 5:00, since EV charging is scheduled. As the purchased PV power is supplied to the base load and EV charging load by the 'PV-system+EV-scheduling' method and the 'PV-system+proposed-EV-scheduling' method, grid power consumption decreases during 8:00-18:00 hours when the PV power is produced. During 19:00-23:00 hours, the grid power is less utilized by the 'no-PV-system+no-EV-scheduling' method as EV charging is executed in the early stages. On the other hand, the 'PV-system+proposed-EV-scheduling' method leads to higher grid power consumption while charging the EVs to target SoC over the time interval 19:00-23:00 hours.

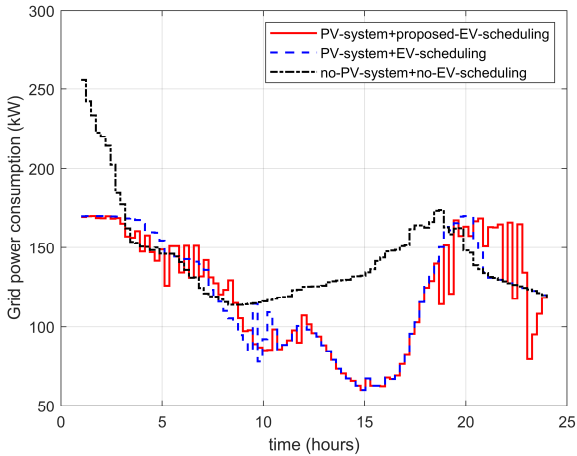

Fig. 6. Variation of grid power consumption.

Figure 7 represents the EV charging/discharging load in the residential parking station. In the 'PV-system+proposed-EV-scheduling' method, EV discharging is executed over the time interval 0:00-5:00 hours to minimize the grid power consumption which can be increased by EV's simultaneous charging load. EV charging is requested over the time interval 18:00-24:00 hours, although the base load is higher than those over other time periods. To mitigate the peak load in this time interval, EV discharging occurs actively. As the 'PV-system+EV-scheduling' method does not include EV discharging, the charging load demand is limited by scheduling. During 0:00-3:00 hours, EV charging load is higher by the 'no-PV-system+no-EV-scheduling' method, as most parked EVs are charged regardless of base load. As the parking station is located in the residential area, EV charging load does not exist between 8:00 and 13:00 hours. The 'PV-system+proposed-EV-scheduling' method reduces the electricity cost by 6.7% and 18.14% compared to the 'PV-system+EV-scheduling' method and the 'no-PV-system+no-EV-scheduling' method, respectively.

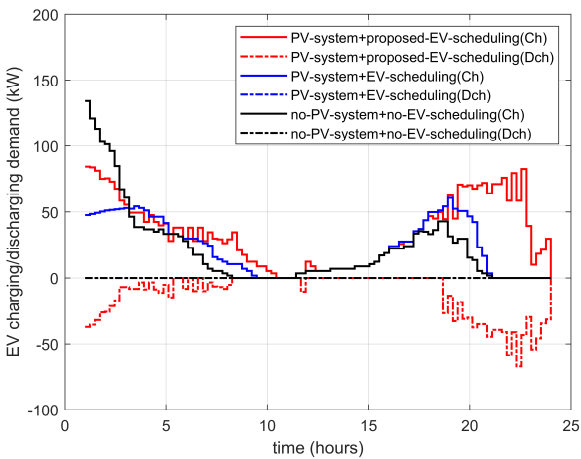

Fig. 7. EVs charging/discharging demand.

## V. CONCLUSION

In this paper, a power management of EV parking station in MG is considered to reduce the electricity cost and grid dependency, and to maximize the utilization of PV power, while taking EVs as ESSs. The MG considered in this paper is characterized by V2G configuration and PV power production. In the proposed scheme, operations of EVs are simultaneously decided depending on the comparison of base load and PV power production. If the temporal base load and PV power production correspond to the EV discharging condition, the EVs can be discharged to reduce the electricity cost and grid power consumption. It is shown that the electricity cost is reduced through EV discharging during the high parking rates, and the PV power supplements the base load in MG, decreasing the peak load in time intervals of low parking rates. Since the time zone in which EV charging is preferred is distinct from the time zone in which EV discharging is preferred, EV charging/discharging is also affected by the base load and PV power production. In simulations, the proposed power management scheme is compared in terms of grid power consumption and total electricity cost. The results show that the proposed power management scheme significantly decreases the total electricity cost and peak load, as compared to the other methods.